\documentstyle[preprint,epsf,aps]{revtex}
\begin{document}
\title{ The Innsbruck quantum teleportation experimental scheme can be modified to the unconditional one
}
\author{Wang Xiang-bin\thanks{email: wang$@$qci.jst.go.jp} 
\\
        Imai Quantum Computation and Information project, ERATO, Japan Sci. and Tech. Corp.\\
Daini Hongo White Bldg. 201, 5-28-3, Hongo, Bunkyo, Tokyo 113-0033, Japan}

\maketitle 
\begin{abstract}
We give a simple way to detersministically 
rule out the event that two pairs of photons are generated in the 
same side of the nolinear crystal in the type II parametric downconversion. By this new 
scheme, everytime when the concindence is observed it is indeed an event that one pair
of photons is generated in each side of the crystal  therefore
 the Innsbruck quantum teleportation experiment(Bouwmeester D et al, Nature 390, 575(1997)) 
can be modified intot the unconditional one.
\end{abstract}
As it is well known that\cite{bennett}, if two remote parties, Alice and Bob share an 
entangled state e.g.
\begin{eqnarray}
|\Psi^-\rangle =\frac{1}{\sqrt 2}(|H\rangle_2|V\rangle_3-|V\rangle_2|H\rangle_3)
\end{eqnarray} 
where $H$ and $V$ are for the horizontal polarizotion and vertical polarization respectively,
Alice may teleport an unknown state $|\chi\rangle_1$ to Bob by making a joint meassurement to
mode 2 and 3 in the 
Bell basis.
Here the subscripts are used to label each photons. The 4 states for the Bell basis are
\begin{eqnarray}
|\Psi^{\pm}\rangle =\frac{1}{\sqrt 2}(|H\rangle|V\rangle\pm|V\rangle|H\rangle)
\end{eqnarray}
and
\begin{eqnarray}
|\Phi^{\pm}\rangle =\frac{1}{\sqrt 2}(|H\rangle|H\rangle\pm|V\rangle|V\rangle).
\end{eqnarray}
To let Bob recover the state $|\chi\rangle$ on particle 3, 
Alice broadcasts her measurement results and Bob will then take a local unitary transformation
to particle 3 according to Alice's result. The first experimental test of the quantum 
teleportation is done by the Innsbruck group\cite{bou}. In the experiment, both the entangled
 state $|\Psi^-\rangle$ and 
 the teleported state $|\chi\rangle$ are produced by the type II parametric downconversion.
As it was commented in\cite{kimble0}, in that set-up there is 
a comparable probability of generating
two pairs in one side of the nonlinear crystal
when the pump light is reflected back while generating nothing in the otherside. 
Taking this factor into
consideration, though it has nontrivially verified the fact of quantum teleportation,
the Innsbruck experimental 
set-up cannot be really used as an unconditional quantum teleportation machine
to teleport an unknown state,
since the average fidelity there does not exceed the threshold of 3/4 
without a destroying post selection.
It was then argued\cite{reply} that the issue can be {\it in principle} solved given a very good
photon detector which distinguishes  one photon and two photons. However, such a very good
photon detector is not possiblly available by our current technology.
To overcome this technical difficulty, one may immediately consider using the cascaded
detection\cite{braun}. As it was investigated in ref\cite{braun}, to keep the normal 
experimental efficiency, the cascaded detection requires the photon detector to detect
one photon sucessfully in a rate higher than $98\%$ . This is obviously not possible with
the current technology. Here we show that actually the problem can be solved in a very simple and 
very feasible way. 

Let us first consider the well-known properties of a polarizing beam splitter(PBS). As it is shown in Fig.1, a
PBS reflects a horizontally polarized photon and transmmits a vertically polarized photon. We can use this 
property to filter one term in the two pair state. We use a scheme as shown in Fig. 2.
We require the following concidence for a sucessful quantum teleportation:\\
1. Both the  detectors of D$_1$ and D$_2$ must be fired.
\\2.  One and only one in D$_3$ and D$_4$ is fired.

Let's first consider the consequence of the second item in the concidence. 
Suppose after it is pumped, the crystal actually generates two pairs in the left side and nothing 
in the right side.
As it was shown in 
ref.\cite{bou1}, the state for the two pair term can be explicitly written in the following form
\begin{eqnarray}
|\psi\rangle_{14}
=\frac{1}{\sqrt 3}(|2H\rangle_1|2V\rangle_4-|HV\rangle_1|HV\rangle_4+|2V\rangle_1|2H\rangle_4)
.\label{two}\end{eqnarray}
In such a case, beam 4 must include two photons. 
The middle term in the right side of eq.(\ref{two}), i.e., the term with one horizontal 
and one vertical photon in beam 4 is immediately ruled out by the second item in our definition of 
coincidence. Because if this term works, there must be one photon in each side of the PBS therefore
both D$_3$ and D$_4$ will be fired, instead of {\it only one} of them 
being fired. Furthermore, lets consider the event that one detector in D$_3$ and D$_4$ is fired. 
For simplicity, we assume that D$_3$
is fired and D$_4$ is silent. We still suppose that actually there are two pairs generated
in the left side of the crystal and nothing in the right side. 
In such a case, since we have already observed that only D$_3$ is fired, beam 4 must include 2 {\it horizontal} photons only.
According to eq.(\ref{two}), beam 1 must include 2 {\it vertical} photons only.
 Now let's  see what happens after beam 1 reaches the beam splitter. A beam splitter will never
change the polarization of these two photons. Since we have placed a horizontal polarizer before
 detector
D$_1$ and a vertical polarizer before D$_2$, therefore in the case that beam 1 includes two vertical 
photons, the detector D$_1$ will be $never$ fired. Consequently, the event is
abandoned by the item 1 of our definition of the concidence.
Similarly, if  D$_4$ is
fired  and D$_3$ silent, the detector D$_2$ will never be fired. Again, the event is abandoned    
by the item 1 of our definition of the concidence.
So far we have seen that  our scheme can indeed rule out all three terms in the right side of
eq.(\ref{two}). That is to say, we can safely filter all the bad events that two pairs are generated
in the left side of the crystal and nothing generated in the right side. 
However, a good event will still have a  chance of $50\%$ probability to survive.  The good event is that 
one pair of photons in each side of the crystal is generated and beam 1 and beam 2 are collapsed to the
state $|\Psi^-\rangle$ after they reach the beam splitter. In such a case, we have half a probability that both D$_1$ and D$_2$ are fired.   
As it has already shown that, in allthe other collapsing for beam 1 and beam 2, at most one detector
in D$_1$ and D$_2$ is fired. Therefore all other collapsing events are ruled out and we draw the conclusion
that, whenever we observed a concindence, the state in beam 1 has been teleported to beam 3 
sucessfully.

In summary, we have given a simple way to make an unconditional quantum teleportation. Our scheme is
just a slight modification of the Innsbruck\cite{bou,pant} scheme which has been experimentally tested already.
 We believe our schme can be carried out easily
based on the current technology.  Our scheme can be used to teleport an arbitrary linear superposition
state of $|H\rangle$ and $|V\rangle$ by just take an appropriate  rotation operation on beam 1 and beam
4. \\
{\bf Acknowledgement:} 
I thank Prof Imai H for support. I thank Dr Matsumoto K, Dr. Tomita A and Prof Wu LA(China science academy)
for useful discussions. 

\begin{figure}
\begin{center}
\epsffile{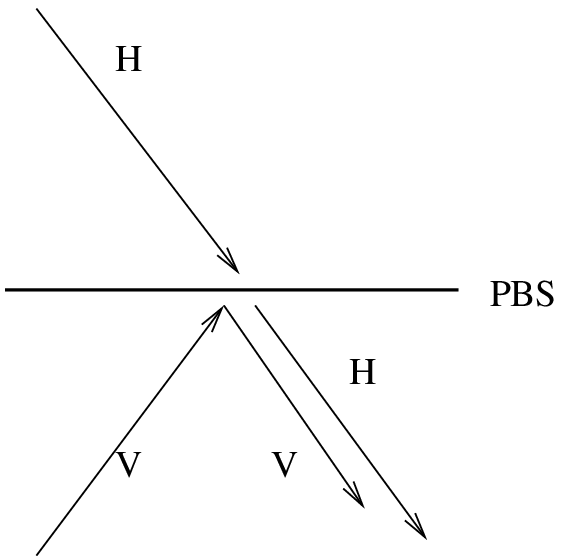}
\end{center}
\caption{A schematic diagram for the properties of a polarizing beam splitter(PBS). It reflects a
vertically polarized photon(V) and transmmits a horizontally polarized photon (H).}
\end{figure}
\newpage
\begin{figure}
\begin{center}
\epsffile{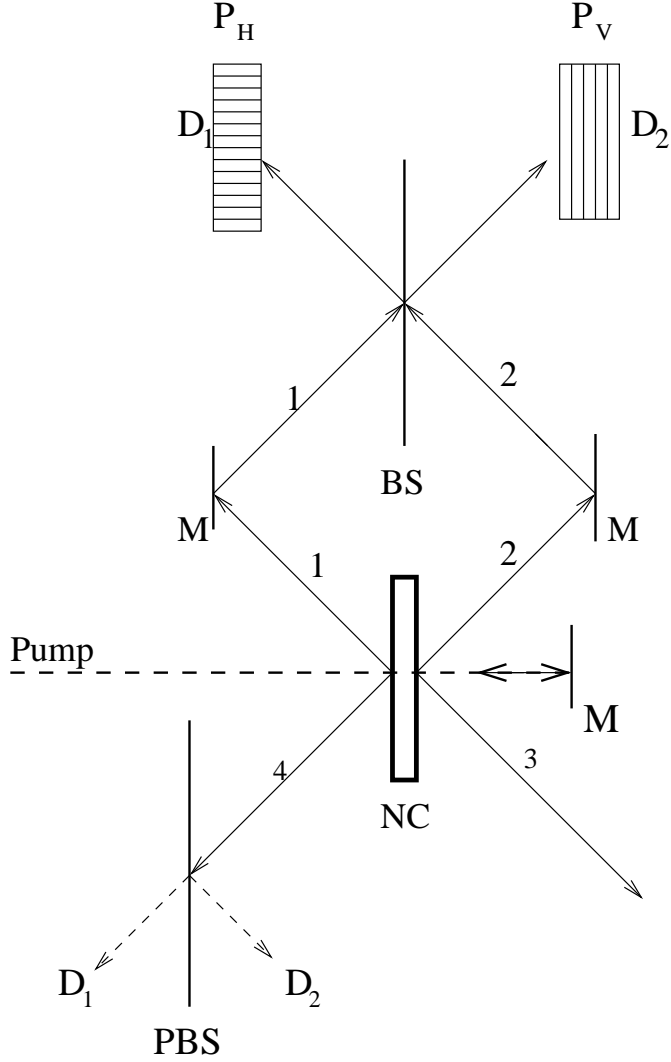}
\end{center}
\caption{A schematic diagram for an unconditional
 quantum teleportation set-up. NC represents for the nonlinear crystal
used in the type II parametric downconversion. BS indicates a beam spliter. PBS is a polarizing beam splitter.
D$_i$ is the $i'$th photon detector. P$_H$ and P$_V$ are horizontal polarizer
and vertical polarizer respectively. Whenever we find the concidence that both D$_1$ and D$_2$ are fired
and one and only one detector in D$_3$ and D$_4$ is fired, the state in beam 1 is teleported
to beam 3. This set-up can be used to teleport an arbitrary unknown state in the linear 
superposition of $|H\rangle$ and $|V\rangle$. D$_3$ and D$_4$ plays a role to both partially rule out 
the satte defined in eq.(\ref{two}) and prepare the state for beam 1.}
\end{figure}
\end{document}